\newcommand{\vdate}{June 1995}
\newcommand{\cernnr}{95-149}
\newcommand{\desynr}{95-107}
\newcommand{\beq}{\begin{equation}}
\newcommand{\eeq}{\end{equation}}
\newcommand{\beqn}{\begin{eqnarray}}
\newcommand{\eeqn}{\end{eqnarray}}

\newcommand{\nonu}{\nonumber\\}

\newcommand{\dd}{\mbox{d}}

\newcommand{\dgsm}[1]{\noindent{\Large\bf #1}}

\documentstyle[12pt,epsfig]{article}

\newlength{\sectionnumbersize}
\newlength{\sectionsize}
\begin{document}

\begin{titlepage}

\renewcommand{\thefootnote}{\fnsymbol{footnote}}
\setcounter{footnote}{0}

\begin{flushright}
\hfill CERN-TH/\cernnr\\
\hfill DESY \desynr\\
\end{flushright}

\vspace{0.5cm}

\begin{center}

{\Large\bf
   {The Mellin Transform Technique\\
    for the Extraction of the Gluon Density\\
   }
}

\vspace{1cm}

\vspace{0.5cm}
\vspace{0.5cm}

{\bf D.\ Graudenz$^1$,
M.\ Hampel$^2$,
A.\ Vogt$^3\footnote{
On leave of absence from Sektion Physik, Universit\"at
M\"unchen, D-80333 Munich, Germany.
}$,
Ch.\ Berger$^2$
}

\vspace{2mm}

\vspace{3mm}
{{}$^1 $Theoretical Physics Division, CERN\\ CH-1211 Geneva 23, Switzerland}

\vspace{3mm}
{{}$^2 $I.\ Physikalisches Institut, RWTH Aachen\\
D-52056 Aachen, Germany}

\vspace{3mm}
{{}$^3 $Deutsches Elektronen-Synchrotron DESY\\ D-22603 Hamburg, Germany}

\vspace{1.0cm}

\footnotetext[0]{Electronic mail addresses:
Dirk.Graudenz@cern.ch, Hampel@desy.de,
avogt@x4u2.desy.de,\\ Berger@rwth-aachen.de.}

\begin{abstract}
A new method is presented to determine the gluon density in the proton from
jet production in deeply inelastic scattering. By using the technique of
Mellin transforms not only for the solution of the scale evolution equation
of the parton densities but also for the evaluation of scattering cross
sections, the gluon density can be extracted in next-to-leading order QCD.
The method described in this paper is, however, more general, and can be used
in situations where a repeated fast numerical evaluation of scattering cross
sections for varying parton distribution functions is required.
\end{abstract}

\end{center}

\vfill
\noindent
\begin{minipage}[t]{5cm}
CERN-TH/\cernnr\\
\vdate
\end{minipage}
\vspace{1cm}
\end{titlepage}

\renewcommand{\thefootnote}{\arabic{footnote}}
\setcounter{footnote}{0}

\newpage
\section{Introduction}
One of the main goals of experiments at the electron--proton
collider HERA is the precise determination of
the gluon density $f_{g/p}(\xi,\mu^2)$ in the proton for various gluon
momentum fractions $\xi$ and factorization scales $\mu$.
In addition to the indirect method
of extracting $f_{g/p}$ from the scaling violation of the structure
function $F_2$,
direct methods such as heavy quark
and jet production have been studied.

In the QCD--improved parton model, the
electron--proton scattering cross section
$\sigma$ is generically given by
a convolution of process-independent parton densities $f_{q/p}$
for (anti-)quarks  and $f_{g/p}$ for gluons with corresponding
mass-factorized
parton-level cross sections $\sigma_q$ and $\sigma_g$:
\beq
\label{fact}
\sigma\,=\,\int \dd\xi\left[f_{q/p}(\xi,\mu^2)\,
\sigma_q(\xi,\mu^2)
+ f_{g/p}(\xi,\mu^2) \sigma_g(\xi,\mu^2)\right].
\eeq
Besides the indicated dependence on $\xi$ and $\mu$,
$\sigma_q$ and $\sigma_g$ also depend on other variables
such as the absolute electron four--momentum transfer squared $Q^2$
and the momenta of the outgoing partons\footnote{Here and in the
following we do not explicitly display the dependence on the renormalization
scale.}.

\begin{figure}[htbp] \unitlength 1mm
\begin{center}
\begin{picture}(160,45)
\put(-0.3,0){\epsfig{file=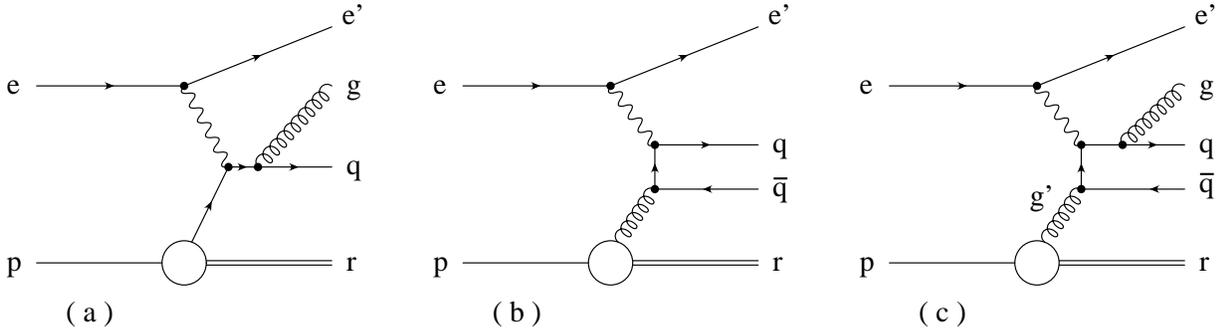,width=16cm}}
\end{picture}
\end{center}
\caption[]
{{\it Generic Feynman diagrams for the leading-order processes of
QCD Compton scattering (a) and photon--gluon fusion (b),
and an example for a diagram corresponding to a next-to-leading
order real correction (c).}}
\label{figref1}
\end{figure}

In leading order (LO), the prescription for the extraction of $f_{g/p}$ from
jet cross sections in deeply inelastic scattering reactions is
very intuitive, because
$\sigma_q$ can be identified with the
parton-model cross section $\sigma_C$
for the so-called {\em QCD Compton scattering} process (fig.~\ref{figref1}a),
and
$\sigma_g$ with the parton-model cross section $\sigma_F$ for the
{\em photon--gluon fusion} reaction (fig.~\ref{figref1}b).
Explicit expressions can be found in
\cite{1}.
Experimentally the outgoing partons from the hard scattering
reactions are identified with jets. The QCD Compton scattering and
photon--gluon
fusion reactions lead to (2+1)-jet final states, where the notation
accounts for the two outgoing jets from the hard scattering process and the
jet in the proton fragmentation region.
The calculated contribution
$\sigma_{C,2+1}^{\mbox{\scriptsize LO}}\,
=\,\int \dd\xi\,f_{q/p}(\xi,\mu^2)\,
\sigma_{C}(\xi)$
{}from Compton scattering can be subtracted from the measured
cross section
\beq
\label{slo}
\sigma_{2+1}^{\mbox{\scriptsize LO}}\,
=\,\int \dd\xi\left[f_{q/p}(\xi,\mu^2)\,
\sigma_{C}(\xi)
+ f_{g/p}(\xi,\mu) \sigma_{F}(\xi)\right],
\eeq
and thus $f_{g/p}(\xi,\mu^2)$ can be determined
in LO by a direct unfolding, since in this case
$\xi$ can be expressed in terms of
measurable quantities as
$\xi=x_B\,\left(1+\hat{s}/Q^2\right)$,
where $x_B$ is the Bjorken scaling variable and $\hat{s}$ is the invariant
mass squared of the system of the two current jets.
An analysis based
on this principle has recently been presented by the
H1 Collaboration \cite{2}.

In next-to-leading order (NLO) this simple picture is destroyed.
Aside from the virtual corrections to the Born processes
in figs.~\ref{figref1}a and b, real corrections have to be
added; diagrams of the type shown in fig.~\ref{figref1}c
can also lead to (2+1)-jet configurations:
if the gluon $g$ attached to the outgoing quark is soft or collinear
to the quark, the diagram constitutes a correction to the photon--gluon
fusion process. If, on the other hand, this gluon is hard and the
outgoing antiquark $\overline{q}$ is soft or collinear to the incoming gluon
$g^\prime$, then this configuration can be said to be a correction
to the QCD Compton scattering reaction.
In the latter case, the collinear or soft antiquark forms a jet with the
proton remnant $r$, and the cross section has to be integrated
over all momenta of the antiquark
according to a specific jet definition scheme.
Collinear singularities
that do not cancel against corresponding singularities from the
virtual corrections
have to be absorbed into renormalized parton densities.
Depending on
the factorization scheme chosen, finite subtracted pieces remain.
The factorization theorems of perturbative QCD
guarantee that the cross section can be written in the form of
eq.~(\ref{fact}).
However, beyond the leading order, the arbitrary momentum
of collinear partons renders the variable $\xi$ unobservable,
because the mass-factorized parton-level cross sections
are in general distributions, not regular functions,
and the simple and straightforward method described above can therefore
not be applied.
A physical consequence is that the distinction between the QCD Compton
and photon--gluon fusion processes becomes meaningless.
Related to this is the fact that quark and gluon densities
mix in the Altarelli--Parisi scale evolution.

A determination of the gluon density in
NLO is very desirable.
In LO the partonic cross sections $\sigma_q$ and
$\sigma_g$ (in short denoted by $\sigma_i$) do not depend on $\mu$,
as already indicated in eq.~(\ref{slo}),
and the $f_{i/p}$ (as short-hand for $f_{q/p}$ and
$f_{g/p}$) are the solutions of the LO Altarelli--Parisi
evolution equation, where the
leading logarithmic terms in the scale $\mu$ are summed up.
In any finite order of perturbation theory,
the scattering cross section $\sigma$ depends explicitly
on the factorization scale $\mu$, this scale dependence being
due to uncalculated higher-order terms. The scale dependence is particularly
strong in the LO case, because there no compensation can take place
between $f_{i/p}$ and $\sigma_i$. To a great extent this problem is, for many
processes, reduced
in NLO, where explicit terms $\sim \ln \mu^2$
in $\sigma_i$ compensate the $\mu$-dependence of $f_{i/p}$ such that the
variation is of higher order in the strong coupling
constant $\alpha_s$.
For reliable theoretical predictions, a NLO analysis of
scale-dependent quantities is therefore mandatory.

The only way to achieve a direct NLO determination of $f_{g/p}$
is
to parametrize the function $f_{g/p}$ at a given scale $\mu_0$,
to evolve it then to
a value of $\mu$ where the cross section is measured, say $\mu=Q$,
and to fit the parameters of $f_{g/p}$
with respect to suitable infrared safe observables, e.g. the (2+1)-jet
cross section in various bins of
$x_B$.
A severe practical problem is that the cross section $\sigma$ has to be
evaluated repeatedly for every choice of parameters for $f_{g/p}$.
Monte Carlo
methods allow the application of arbitrary
cuts on final-state particle momenta, as is necessary in order to
take detector acceptance cuts properly into account, but these methods are
prohibitively slow. A fast numerical method for
the repeated application of this procedure
is indispensable, and will be developed in
this paper.

The paper is organized as follows. The new method is formally derived
in Section~\ref{nfmtt}. Details of the Mellin transform relevant to the
application of the method are discussed in Section~\ref{vogt}.
Finally, an explicit numerical example is given in Section~\ref{hampel}
for the case of
jet cross sections in deeply inelastic electron--proton scattering,
where it is shown that the method is operational in practice
when realistic acceptance
cuts are taken into account.
The paper closes with a short summary.

\section{The Mellin Transform Technique\newline
for Non-Factorizing Cross Sections}
\label{nfmtt}
\label{dg}
The Mellin transform technique allows for a quick numerical evaluation
of integrals of the form
\beq
\label{conv}
\Sigma(x_B)\,=\,\int_{x_B}^1\,\frac{\dd \xi}{\xi}\,f_{i/p}(\xi)
\,\sigma_i\left(\frac{x_B}{\xi},x_B\right)
\eeq
in the case where $\sigma_i$ is independent of its second argument $x_B$,
on the basis of the moments defined by
\beq
\label{mom}
F_n\,\equiv\,\int_0^1\frac{\dd x}{x}\,x^n\ F(x)
\eeq
for an arbitrary function $F$ and
(complex) $n$. The moments of the function
$\Sigma$ are then given by
\beq
\Sigma_n\,=\,f_{i/p,n}\,\sigma_{i,n}.
\eeq
The functional dependence of
$\Sigma$ can be recovered from the moments $\Sigma_n$
by an inverse Mellin transform.
An expression of the form of eq.~(\ref{conv}) will be called to be of the
{\it factorizable type} if the only dependence on $x_B$ in the arguments
of $\sigma_i$ is via $x_B/\xi$. In the application which we
have in mind, $f_{i/p}$
is a parton density, whereas $\sigma_i$ is an expression
for a mass-factorized parton-level
scattering cross section. In general, acceptance cuts
and non-factorizable jet algorithms (cf.\ the discussion in
\cite{3,4})
introduce
an explicit dependence of $\sigma_i$ on $x_B$.
Moreover, the expression for $\Sigma(x_B)$ is integrated over a certain range
of $x_B$. This might suggest that the Mellin transform technique
cannot be applied. However, this is not the case. In the following we outline
a method that allows the use of this technique.

The cross section differential in $x_B$ can be written in
the form
\beq
\label{cc}
\Sigma(x_B)\,=\,\int_{x_B}^1\,\frac{\dd \xi}{\xi}\,f_{i/p}(\xi,\mu^2)\,
\sigma_i\left(\frac{x_B}{\xi},x_B,\mu^2\right),
\eeq
where
\beq
\label{eq7}
\sigma_i\left(\frac{x_B}{\xi},x_B,\mu^2\right)\,=\,\int_{V_{x_B}}\,\dd T
\,\,\hat{\sigma_i}\left(\frac{x_B}{\xi},x_B,T,\mu^2\right).
\eeq
Here we have made the dependence on the factorization scale $\mu$ explicit.
The set $T$ of variables contains all other
integration variables besides $x_B$, i.e.\ the other electron variables
including $Q^2$ and the momenta of the outgoing partons (or jets);
and $V_{x_B}$ is the phase space region over which
these variables are integrated. Thus, $V_{x_B}$ includes all
acceptance and jet cuts, as well as the range in $Q^2$ for the specified
$x_B$.
$\hat{\sigma_i}$ is the cross section differential in all variables
including $T$, whereas
$\sigma_i$ is the integration kernel to be convoluted with
$f_{i/p}$ to yield $\Sigma$.
The explicit dependence on $x_B$ of
$\hat{\sigma_i}(x_B/\xi,x_B,T,\mu^2)$ and $\sigma_i(x_B/\xi,x_B,\mu^2)$
(not just
via the ratio
$x_B/\xi$) is a consequence of the (finite) factorization breaking
terms\footnote{It should be kept in mind that the infinite terms related
to infrared and collinear singularities still factorize in the
standard form so that universal parton densities can be defined. This
property is not spoiled by acceptance cuts and non-factorizing jet
definition schemes.}.

It is assumed that the
factorization scale $\mu$ is chosen independently from the variables $T$;
in particular, $\mu$ should not depend on the integration variable $Q^2$,
which is integrated over a certain range
$[Q_0^2,Q_1^2]$. $\mu$ it may be set to some intermediate value.
In any case, this is not a  strong restriction, because the parton densities
depend only logarithmically on the factorization scale,
and moreover
the scale dependence is compensated by a corresponding term
in the mass-factorized parton-level scattering cross section,
so that the change is of higher order in
$\alpha_s$.
For simplicity of notation, we drop the argument $\mu^2$ in
the following.

Now let (for a fixed $Q^2$-bin) $a_1$, \ldots, $a_k$ be the
experimental boundaries
of the intervals in the variable $x_B$
for which the cross sections are measured.
To proceed, we define
\beq
\label{ccc}
\Sigma_a\,\equiv\,\int_a^1\,\dd x_B\,\Sigma(x_B).
\eeq
The integral over a specified interval $[a_i,a_{i+1}]$ in $x_B$ is then
simply given by
\beq
\label{dg2}
\label{eq9}
\int_{a_i}^{a_{i+1}}\,\dd x_B\,\Sigma(x_B)=\Sigma_{a_i}-\Sigma_{a_{i+1}}.
\eeq
In a fit of the function $f_{i/p}$,
the integrals in eqs.~(\ref{cc}), (\ref{eq7}), (\ref{ccc})
have to be evaluated repeatedly. In general, the $x_B$, $\xi$ and
$T$-integrations are performed by a time-consuming Monte Carlo integration
in order to implement all cuts. In particular, the $(x_B,Q^2)$-plane
is divided into several bins which do not change during the fitting
procedure.
The method requires that a certain set of moments is calculated for every
bin in the $(x_B,Q^2)$-plane. The steps to be followed to determine
the $\Sigma_a$ are:

\begin{itemize}
\item
Define a function
\beq
h_a(u)\,\equiv\,
\left\{
\begin{array}{cl}
{\displaystyle \int_a^{a/u}\,\dd x_B \,\sigma_i\left(
\frac{x_B}{a/u},x_B\right)},& \mbox{if}\quad u\geq a\\
0,& \mbox{if}\quad u<a\\
\end{array}
\right.
\eeq
and its moments in the variable $u$
\beq
h_{an}\,\equiv\,\int_0^1\,\frac{\dd u}{u}\,u^n\,h_a(u).
\eeq
It is easy to prove that an explicit expression for $h_{an}$
is
\beq
\label{eq12}
\label{dg1}
h_{an}\,=\,\int_a^1\,\dd x_B \,\int_{x_B}^1\,\frac{\dd \xi}{\xi}
\,\left(\frac{a}{\xi}\right)^n\,\sigma_i\left(\frac{x_B}{\xi},x_B\right).
\eeq
The $h_{an}$ are thus the $\Sigma_a$ with the parton density
$f_{i/p}(\xi)$ replaced by $(a/\xi)^n$. They can be determined numerically
by means of a Monte Carlo integration. In general, for complex $n$,
the quantity $(a/\xi)^n$ has to be split into its real and imaginary
part.

\item
Define
\beq
\Sigma_{ab}\,\equiv\,\int_a^1\,\frac{\dd \xi}{\xi}\,
f_{i/p}(\xi)\,h_b\left(\frac{a}{\xi}\right)
\eeq
and determine the moments of $\Sigma_{ab}$ with respect to the variable $a$:
\beq
\tilde{\Sigma}_{nb}\,\equiv\,\int_0^1\,\frac{\dd a}{a}\,
a^n\,\Sigma_{ab}.
\eeq
Obviously, $\Sigma_a=\Sigma_{aa}$.
The key relation of our method is
\beq
\label{central}
\tilde{\Sigma}_{nb}=f_{i/p,n}\,h_{bn},
\eeq
and can be proved in the
following way:
\beqn
\tilde{\Sigma}_{nb}\,
&=&\,
\int_0^1\,\frac{\dd a}{a}\,a^n\,\int_a^1\,
\frac{\dd \xi}{\xi}\,f_{i/p}(\xi)\,h_b\left(\frac{a}{\xi}\right)\nonu
&=&\,
\int_0^1\,\frac{\dd \xi}{\xi}\,f_{i/p}(\xi)\,\int_0^\xi
\,\frac{\dd a }{a}\,a^n\,h_b\left(\frac{a}{\xi}\right)
\nonu
&=&\,
\int_0^1\,\frac{\dd \xi}{\xi}\,\xi^n\,f_{i/p}(\xi)\,\int_0^1\,\frac{\dd u}{u}\,
u^n\,h_b(u)
\nonu
&=&\,
f_{i/p,n}\,h_{bn}.
\eeqn

\item
For a given parametrization of $f_{i/p}$ in terms of its moments
$f_{i/p,n}$ the cross section $\Sigma_a$ can be determined by forming the
moments $\tilde{\Sigma}_{na}=f_{i/p,n}\,h_{an}$ and a subsequent
inverse Mellin transform in the variable $n$, evaluated at $a$.

\end{itemize}

The Mellin transform method in the case
of non-factorizing cross sections introduces the inconvenience that
the moments $h_{an}$ have to be
determined for every interval boundary $a$ separately. Due to the large
number of repeated cross section evaluations in the fitting procedure,
however, this method
is far more efficient than a direct integration of
the integrals in eqs.~(\ref{cc}), (\ref{eq7}), (\ref{ccc})
for every parametrized parton density.

\section{From Parton Moments to Observables}
\label{vogt}
\begin{figure}[htbp] \unitlength 1mm
\begin{center}
\begin{picture}(160,70)
\put(20,-5){\epsfig{file=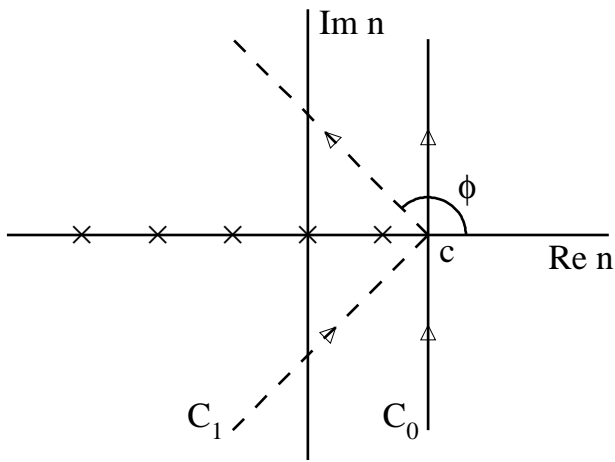,width=10cm}}
\end{picture}
\end{center}
\caption[]
{{\it Integration contours of the Mellin inversion in eq.~(\protect\ref
  {inv}), leading to the inversion formulae of eqs.~(\protect\ref{inv1}) and
 (\protect\ref{inv2}) for the routes ${\cal C}_{0}$ and ${\cal C}_{1}$,
 respectively. The crosses schematically denote the singularities of
 $F_{n}$. }}
\label{avfig1}
\end{figure}

Let us now consider the inverse transformation of the moments given by
eq.~(\ref{mom}), which is a special case of the general Mellin
transformation for functions $F(x)$ vanishing identically at $x>1$.
If $F(x)$ is piecewise smooth for $x>0$, the corresponding Mellin
inversion reads
\beq
\label{inv}
  F(x) = \frac{1}{2\pi i} \int_{c-i\infty}^{c+i\infty} \! \dd n \, x^{-n}
         F_{n} \:\: ,
\eeq
where the real number $c$ has to be chosen such that $ \int_{0}^{1} dx
\, x^{c-1} F(x) $ is absolutely convergent \cite{5}. Hence $c$ has to
lie to the right of the rightmost singularity $n_{max}$ of $F_n$. The
contour of the integration in eq.~(\ref{inv}) is displayed in
fig.~\ref{avfig1} and denoted by ${\cal C}_{0}$. Also shown is a deformed route
${\cal C}_{1}$, yielding the same result as long as no singularities
$n_i$ of $F_{n}$ are enclosed by ${\cal C}_{0} - {\cal C}_{1}$.
For example, for the
LO and NLO evolution of structure functions, the $n_i$ are real
with $ n_i < n_{max} < c $, and this requirement is fulfilled
automatically.

It is useful to rewrite eq.~(\ref{inv}) as an integration over a real
variable. We are concerned with functions obeying $ F^{\ast} _{n} =
F_{n^{\ast}} $, where `$\ast$' denotes the complex conjugation.
Then it is easy to show that eq.~(\ref{inv}) yields,
for the contour
characterized by the abscissa $c$ and the angle $\phi $ in
fig.~\ref{avfig1}:
\beq
\label{inv1}
  F(x) = \frac{1}{\pi} \int_{0}^{\infty} \! \dd z \, \mbox{Im} \left[
         \exp{(i\phi)} \, x^{-c-z\exp{(i\phi)}} F_{n=c+z\exp{(i\phi)}}
         \right].
\eeq
It is obvious from the discussion given above that the
integral does not depend on $c$ and $\phi $. However, for an efficient
numerical evaluation a suitable choice of these parameters is very
useful. For example, it is advantageous to choose $ \phi > \pi /2 $ in case
$F_{n}$ is a known analytical function, especially if this
function does not fall off very rapidly for $|n| \rightarrow \infty $.
The factor $ \exp \left(z \log{\frac{1}{x}} \cos{\phi}\right) $ then
introduces an exponential dampening of the integrand
(which rapidly oscillates at small~$x$)
with increasing $z$, thereby allowing for a
smaller upper limit $z_{max}$ in the numerical implementation of
eq.~(\ref{inv1}). This procedure has been employed for the inversion of
moments of parton densities and structure functions for the proton
and the photon, e.g.\ in \cite{6,7} and \cite{8},
respectively.

In general, however, the moments of the partonic cross section can only
be calculated numerically using eq.~(\ref{mom}), because no analytic
continuation to small Re$\,n$, where the integral does not exist, is at
our disposal. Likewise, in our case these moments are given by
eq.~(\ref{dg1}) and do not behave uniformly for $|n| \rightarrow \infty $.
Especially, they grow exponentially along ${\cal C}_{1}$. Therefore, we
will use the `textbook contour' ${\cal C}_{0}$ in the following and,
with $\phi = \pi /2$, eq.~(\ref{inv1}) simplifies to
\beq
\label{inv2}
  F(x) = \frac{1}{\pi} \int_{0}^{\infty} \! \dd z \, \mbox{Re} \left[
         x^{-c-iz} F_{n=c+iz} \right] \:\: .
\eeq
We have applications in mind where $ F_{n} = f_{i/p,n} h_{an}$, see
eq.~(\ref{central}), and the numerical evaluation of the moments $h_{an}$
in eq.~(\ref{eq12}) is {\it very} time-consuming.
Taking a different upper limit
$z_{max} $ of the numerical $z$-integration or
number of points for the integral at each step in the integration process
is practically unfeasible in such a case. Instead, we want to fix
$z_{max} $ at a value as small as possible in order to allow for an
evaluation of eq.~(\ref{inv2}) with a rather small number of fixed moments.

In this context, it is useful to consider that part of the $n$-dependent
inversion integrand, of which the analytical continuations are known
together with the behaviour at large $n$, namely the parton densities
and their evolution. Inspection of eq.~(\ref{eq12})
suggests that $ h_{an} $ does
not rise strongly with $z$ along ${\cal C}_{0}$. Hence the large-$n$
behaviour of the parton densities $f_{i/p,n}$ can be employed to estimate
the convergence of the complete integral in eq.~(\ref{inv2}) with $F_{n}$ given
by eq.~(\ref{central}). A typical ansatz for the parton distribution
functions of the proton at some reference scale $ \mu^{2}_{0} $, denoted
by $ f_{i/p}(\xi)$, is given by \cite{7,9}
\beq
\label{inp}
  \xi f_{i/p}(\xi) = A\, \xi ^{\alpha}\, (1-\xi )^{\beta}\, (1+\gamma
            \sqrt{\xi}+ \ldots) \:\: .
\eeq
The coefficients $\beta $ can be estimated roughly by their counting
rule values, e.g. $ \beta_{\rm val} \approx 3 $, $\beta_{\rm glue}
\approx 5 $. The Mellin transform of eq.~(\ref{inp}) reads simply
\beq
  f_{i/p,n} = A \left[B(\alpha +n-1,\beta +1) + \gamma B(\alpha +n-1/2,
          \beta +1) + \ldots \right]
\eeq
with the Euler Beta-function $B$. If $\beta $ is a positive
integer, then this equation simplifies to
\beq
  f_{i/p,n} = \frac{A\, \beta !}{(\alpha + n-1)(\alpha + n)
        \ldots (\alpha +n+\beta-1)} + \ldots = {\cal{O}}(1/n^{\beta +1})
        \mbox{ for } n \rightarrow \infty \:\: .
\eeq
The evolution of these input moments is known analytically for arbitrary
complex $n$ \cite{10}, and the kernel $K_{ij,n}(\mu^2, \mu_{0}^2)$ in
\beq
\label{evol}
  f_{i/p,n}(\mu ^2) = K_{ij,n}(\mu^2,\mu_{0}^2)\, f_{j/p,n}(\mu_{0}^2)
\eeq
generally leads to a slightly faster decrease of $ f_{i/p,n}$ for large
$n$ at $\mu^2 > \mu_{0}^2 $. Hence a fall-off like $1/n^4$ can be safely
used to estimate the practically required upper limit in
eq.~(\ref{inv2}).

For this purpose, we have numerically determined the upper limit
$z_{max}$ sufficient to reach a $1\%$ accuracy of the Mellin inversion
for the toy function $ F(x) = x^{-1}(1-x)^3 $. The results are
displayed in table~\ref{tabrefv} for selected values of $x$.
The rightmost pole is at $ n_{max} = 1 $, and we have chosen $c = 1.5$.
The larger $c$ is, the more $F_{n=c+iz}$ is flattened. Hence too large a
value of $c$ leads to an undesired rise of $z_{max}$ at small $x$,
where $F_{n}$ is integrated after multiplication with a rapidly
oscillating function in eq.~(\ref{inv2}). Practically, $ c - n_{max}
\simeq 0.5 $ works well for all $x$-values of interest here, implying
$ c \simeq 1.8 $ for realistic small-$\xi $ parton densities
\cite{7,9}.

\begin{table}[htb]
  \begin{center}
  \begin{tabular}{|c||c|c|c|c|c|c|c|}\hline
           &           &           &      &      &      &      &
                                                        \\[-0.4cm]
  $x$      & $10^{-4}$ & $10^{-3}$ & 0.01 & 0.03 & 0.1  & 0.3  & 0.8
                                                        \\[0.1cm]\hline
           &           &           &      &      &      &      &
                                                        \\[-0.4cm]
  $z_{max}$&   6.0     &  5.5      & 4.5  & 4.0  & 3.5  & 5.0  & 8.0
                                                        \\[0.1cm]\hline
  \end{tabular}
  \caption{\label{tabrefv}
{\it Upper limits $z_{max}$ numerically sufficient for a $1\%$
   accuracy of the Mellin inversion from eq.~(\protect\ref{inv2})
   for the function $ F(x) = x^{-1}(1-x)^3 $  with $ c=1.5 $.}}
  \end{center}
\end{table}

The integral eq.~(\ref{inv2}), truncated at $z_{max}$, can now be
performed by using a sufficiently large number of fixed support points,
e.g. by a sum of 8-point Gaussian quadratures, see \cite{11} for the
weights and support points. In this way, everything except for the
input-$B$-functions varied in a fit of the parton distribution functions
is fixed; especially the time-consuming part of the kernels in
eq.~(\ref{evol}) and the moments $ h_{an} $ of the partonic cross sections
{}from eq.~(\ref{dg1}) can be determined once and then used unchanged in the
calculation of physical observables for various parton densities.

\section{Application to Jet Physics at HERA}
\label{hampel}
To illustrate how the Mellin transform method can be used to fit the gluon
density $f_{g/p}$,
the gluon-induced (2+1)-jet cross sections were calculated in several bins
for HERA energies of $820\, \mbox{GeV}$ protons and
$27.6\, \mbox{GeV}$ electrons.
Quark contributions were set to zero explicitly in the parton
distribution function
to reduce the number of moments needed for this case study.

The program PROJET~\cite{12} based on the
NLO matrix elements from \cite{13,14} was used, this
allows to calculate jet cross sections in LO and NLO in
the modified JADE scheme
defined in the following way
\cite{15}:

\begin{itemize}
\item Define a {\it precluster} of longitudinal momentum $p_r$
given by the missing longitudinal momentum of the event.
\item Apply the JADE cluster algorithm \cite{16} to the set of momenta
$\{p_1,\ldots,p_n,p_r\}$, where $p_1,\ldots,p_n$ are the momenta
of the hadrons visible in the detector. The resolution criterion
is $s_{ij}=2p_ip_j>y_{cut}M^2$.
Here $M^2$ is a mass scale and
$y_{cut}$ is the resolution parameter.
\end{itemize}
In the case of a theoretical calculation, $p_r$ is directly given by
the momentum fraction of the proton not carried by the incident parton,
and $p_1,\ldots,p_n$ are the momenta of the partons in the final state.
In the following, we choose $W^2$, the squared total hadronic energy,
as the mass scale $M^2$, since the proton remnant is included in the
jet definition.

The integration routine used in PROJET is VEGAS~\cite{17,18}.
As is desirable for an experimental measurement, the phase space was
binned in~$Q^2$ and~$x_B$ according to
eq.~(\ref{ccc}); the bins are given
in tables~\ref{complo} and~\ref{compnlo}.
In addition, the following typical H1 detector cuts were applied,
for which the motivation is explained in detail in~\cite{19}:

\begin{itemize}
\item{The invariant mass squared of the hadronic system $W^2$ was
      required to be larger than $5000\, \mbox{GeV}^2$.}
\item{The jet resolution cut $y_{cut}$ was set to $0.02$.
      Lowering this value significantly below $0.01$ causes NLO corrections
      to dominate and leads to
      unphysical cross sections. It is important to note that
      $\xi \geq y_{cut}$ as a consequence of the applied
      modified JADE algorithm. The region $\xi >0.01$ is however
      very interesting~\cite{20} for a precise
      determination of $f_{g/p}$, see also \cite{21}.}
\item{The jets were required to lie in the polar angle range
      of $10^{\circ} \leq \theta_{jet} \leq 145^{\circ}$.}
\item{For bins with $Q^2 \leq 100\, \mbox{GeV}^2$,
      the scattered electron had to have
      an energy of $E_{l'} \geq 14\, \mbox{GeV}$ and the polar angle had to
      lie within
      the range of $160^{\circ} \leq \theta_{l'} \leq 172.5^{\circ}$.}
\item{In the bins with $Q^2 \geq 100\, \mbox{GeV}^2$,
      the scaled photon energy~$y$ in the proton rest system
      had to be $y \leq 0.7$ and
      the scattered electron was required to have
      $10^{\circ} \leq \theta_{l'} \leq 148^{\circ}$.}
\end{itemize}
In this list, angles and energies
are defined in the laboratory frame, and angles are given
with respect to the direction of the incoming proton.
For each bin, 32 complex Mellin moments were calculated according to the
prescription described in Section~\ref{nfmtt}, cf.\ eq.~(\ref{dg1}).
In all calculations, $\alpha_s$ was computed to second order, and
the NLO gluon distribution function of~\cite{7}
was employed.

\addtocounter{footnote}{1}
\newcounter{bergerlabel}
\setcounter{bergerlabel}{\value{footnote}}

\begin{table}[htb]
\centering
\begin{tabular}{|rcr||c|c||c|c||c|c||c|c||}
\cline{4-11}
\multicolumn{3}{c}
{~} &
\multicolumn{8}{|c|}
{
$x_B$
} \\
\hline
\multicolumn{3}{|c||}
{
$Q^2$ [$\mbox{GeV}^2$]
}
&
\multicolumn{2}{c||}{$10^{-4}\ldots1$} &
\multicolumn{2}{c||}{$10^{-3}\ldots1$} &
\multicolumn{2}{c||}{$10^{-2}\ldots1$} &
\multicolumn{2}{c|}{$10^{-1}\ldots1$} \\
\hline
\hline
$10$&\ldots&$14$ & $62.80$ & $61.64$ & $28.09$ & $28.74$
& --- & --- & --- & --- \\
\hline
$14$&\ldots&$18$ & $70.64$ & $69.72$ & $50.95$ & $49.97$
& --- & --- & --- & --- \\
\hline
$18$&\ldots&$25$ & $85.82$ & $84.89$ & $71.03$ & $69.80$
& --- & --- & --- & --- \\
\hline
$25$&\ldots&$40$ & $109.9$ & $108.8$ & $101.9$ & $101.1$
& --- & --- & --- & --- \\
\hline
$40$&\ldots&$100$ & --- & ---
& $123.8$ & $124.4$ & $14.51$ & $14.43$ & --- & --- \\
\hline
$100$&\ldots&$300$ & --- & ---
& $31.96$ & $32.18$ & $14.69$ & $14.76$ & --- & --- \\
\hline
$300$&\ldots&$700$ & --- & ---
& $28.97$ & $29.23$ & $25.18$ & $25.42$ & --- & --- \\
\hline
$700$&\ldots&$4000$ & --- & ---
& --- & --- & $10.22$ & $10.12$ & $0.96$ & $0.93$ \\
\hline
\end{tabular}
\caption[]{{\it Comparison of cross sections with LO
         matrix elements$\!\;^{\mbox{\scriptsize\arabic{bergerlabel}}}$
         (in [$pb$]) obtained
         by integrating directly (left columns) or using the
         Mellin transform method
         (right columns).}}
\label{complo}

\end{table}

\begin{table}[htb]
\centering
\begin{tabular}{|rcr||c|c||c|c||c|c||c|c||}
\cline{4-11}
\multicolumn{3}{c}
{~} &
\multicolumn{8}{|c|}
{
$x_B$
} \\
\hline
\multicolumn{3}{|c||}
{
$Q^2$ [$\mbox{GeV}^2$]
}
&
\multicolumn{2}{c||}{$10^{-4}\ldots1$} &
\multicolumn{2}{c||}{$10^{-3}\ldots1$} &
\multicolumn{2}{c||}{$10^{-2}\ldots1$} &
\multicolumn{2}{c|}{$10^{-1}\ldots1$} \\
\hline
\hline
$10$&\ldots&$14$ & $58.48$ & $57.25$ & $26.60$ & $26.00$
& --- & --- & --- & --- \\
\hline
$14$&\ldots&$18$ & $66.57$ & $65.90$ & $47.22$ & $46.69$
& --- & --- & --- & --- \\
\hline
$18$&\ldots&$25$ & $82.48$ & $81.65$ & $67.99$ & $66.87$
& --- & --- & --- & --- \\
\hline
$25$&\ldots&$40$ & $108.1$ & $107.4$ & $100.4$ & $99.71$
& --- & --- & --- & --- \\
\hline
$40$&\ldots&$100$ & --- & ---
& $126.1$ & $125.6$ & $14.07$ & $13.96$ & --- & --- \\
\hline
$100$&\ldots&$300$ & --- & ---
& $34.86$ & $34.52$ & $15.51$ & $15.31$ & --- & --- \\
\hline
$300$&\ldots&$700$ & --- & ---
& $31.34$ & $31.51$ & $27.01$ & $27.19$ & --- & --- \\
\hline
$700$&\ldots&$4000$ & --- & ---
& --- & --- & $11.18$ & $11.19$ & $0.99$ & $0.97$ \\
\hline
\end{tabular}
\caption{{\it Comparison of NLO cross sections (in [$pb$]) obtained
         by integrating directly (left columns) or using the
         Mellin transform method
         (right columns).}}
\label{compnlo}
\end{table}

A good convergence of the
numerical calculations was found for
$c = 1.8$,
$\phi = \pi/2$ and $z_{max} = 9$, with a higher density of
support points at lower $z$,
as the influence is greatest there.
For comparison, the cross section was also calculated directly, see
eqs.~(\ref{cc}), (\ref{eq7}), (\ref{ccc}).
After inverting the
product of the hard subprocess and evolved gluon density
moments at the average $Q^2$, the results were found to coincide at the
per cent level.
The detailed results can be found in tables~\ref{complo} and~\ref{compnlo}.
In most bins, convergence was reached at~$z_{max}=3$ (corresponding to
16 moments), the additional moments were used for safety.
The convergence of the LO cross section was much faster than in
the NLO case, as for a given number of support points in VEGAS, the LO
integration is more accurate due to the simpler integration kernel.
The method works well for both LO and NLO.

\footnotetext{Here,
`LO' means that the matrix elements were calculated in LO,
but $\alpha_s$ and the parton distribution functions in NLO
to facilitate a comparison with the results of table~\ref{compnlo}.
For a physically meaningful comparison of the LO with the NLO,
$\alpha_s$ and the parton distribution functions should be calculated
in LO, if they are used in conjunction with the LO matrix elements.}

The number of points in the Monte Carlo integration
was chosen such that the error returned by VEGAS was
less than~$1\%$.
This number is,
however, only a rough estimate~\cite{17,18}, and the achieved accuracy
was studied by
repeating the calculation for different random number generator seeds.
The direct integrations performed here
had a statistical variation of $2$--$3\%$.
The partonic cross section
{}from the Mellin transform method is implicitly integrated
repeatedly by the calculation of the moments, which
smoothes out statistical variations. The results were found to be more
stable than the direct integration, which varied around the result obtained
by the moment inversion.
Even drastic errors of single moments or setting single moments to zero
could be tolerated and led to a reproducible result.
We conclude that this method is numerically very stable and
that the accuracy is of the order of 1\%. Increasing the accuracy
requires increasing the number
of support points for the integration, which
would result in a dramatic increase in CPU
time\footnote{The numerical calculation of one moment needed
about $2$ minutes of CPU time on an SGI Challenge processor in LO, and
about $20$ minutes in NLO.}.
One has to keep in mind that an additional error source arises from
the Mellin transform method,
as for each experimental bin in $x$ one has to calculate the difference of the
cross sections depending on the bin
boundaries in
eq.~(\ref{eq9}),
leading to error propagation.
A strategy for bin optimization is under study.

\section{Summary}
We have outlined a new method, based on the Mellin transform
for the fast evaluation of the convolution
of a parton-level cross section with a parton density, which works
for cross sections not showing a simple factorization behaviour.
The method can be the basis for a fit of the gluon density from
experimental data for the (2+1)-jet cross section
in deeply inelastic electron--proton
scattering, but it is also suitable for more complicated observables.
It has been explicitly shown by a numerical study using
realistic experimental cuts, that the method works in practice with
a sufficiently high accuracy.

\vspace{5mm}
\dgsm{Acknowledgements}

\noindent
It is a pleasure
to thank G.~Kramer for a critical reading of the manuscript
and R.~Nisius for many helpful discussions.
This work was supported in part by the Bundesministerium
f\"ur Bildung und Forschung.
M.H. gratefully acknowledges support by the
Studienstiftung des deutschen Volkes.


\newcommand{\scs}{\rm}
\newcommand{\bibitema}[1]{\bibitem[#1]{#1}}
\newcommand{\bibbeginlong}{\begin{thebibliography}{XXX\,1988\,\,}}
\newcommand{\bibbeginshort}{\begin{thebibliography}{XX}}

\bibbeginshort 

\addcontentsline{toc}{section}{References}

\bibitema{1}
       A.~Mendez, Nucl.\ Phys.\ \underline{B145} (1978) 199.

\bibitema{2}
      H1 Collaboration, S.~Aid et al.,
      preprint DESY 95-086 (1995).

\bibitema{3}
      S.~Catani, Y.L.~Dokshitzer and B.R.~Webber,
      Phys.\ Lett.\ \underline{B285} (1992) 291.

\bibitema{4}
      B.R.~Webber, J.\ Phys.\ \underline{G19} (1993) 1567.

\bibitema{5}
      R.~Courant and D.~Hilbert, {\it Methoden der Mathematischen Physik},
      Springer Verlag, Berlin, 1924.

\bibitema{6}
      M.~Gl\"uck, E.~Reya and A.~Vogt,
      Z.\ Phys.\ \underline{C53} (1992) 127.

\bibitema{7}
      M.~Gl\"uck, E.~Reya and A.~Vogt, preprint DESY 94-206 (1994)
      and Univ.\ Dortmund preprint
      DO-TH 94/24, to appear in
      Z.\ Phys.\ \underline{C}.

\bibitema{8}
      M.~Gl\"uck, E.~Reya and A.~Vogt,
      Phys.\ Rev.\ \underline{D45} (1992) 3968,
                   \underline{D46} (1993) 1973.

\bibitema{9}  A.D.~Martin, W.J.~Stirling and R.G.~Roberts,
                   Phys.\ Rev.\ \underline{D50} (1994) 6734, Rutherford
                   Appleton Laboratory preprint RAL-95-021 (1995).

\bibitema{10}
      M.~Gl\"uck, E.~Reya and A.~Vogt,
      Z.\ Phys.\ \underline{C48} (1990) 471.

\bibitema{11}
      M.~Abramowitz and I.A.~Stegun (eds.), {\it Handbook of
      Mathematical Functions}, National Bureau of Standards, 1964.

\bibitema{12}
      D.~Graudenz, PROJET 4.13 manual, preprint CERN-TH.7420/94
      (November 1994).

\bibitema{13}
      D.~Graudenz,
      Phys.\ Lett.\ \underline{B256} (1991) 518.

\bibitema{14}
      D.~Graudenz, Phys.\ Rev.\ \underline{D49} (1994) 3291.

\bibitema{15}
      D.~Graudenz and N.~Magnussen,
      in: Proceedings of the HERA Workshop 1991,
      DESY (eds.\ W.~Buchm\"uller, G.~Ingelman).

\bibitema{16}
      JADE Collaboration, W.~Bartel et al.,
      Z.\ Phys.\ \underline{C33} (1986) 23.

\bibitema{17}
      G.~Lepage, J.\ Comput.\ Phys.\ \underline{27} (1978) 1992.

\bibitema{18}
      G.~Lepage, Cornell preprint CLNS-80/447 (1980).

\bibitema{19}
      H1 Collaboration, T.~Ahmed et al.,
      Phys.\ Lett.\ \underline{B346} (1995) 415.

\bibitema{20} W.J.~Stirling, talk presented on the
                    Workshop on Deep Inelastic
                    Scattering and QCD, Paris, April 1995.

\bibitema{21}
   W.~Vogelsang and A.~Vogt,
   Rutherford Appleton Laboratory preprint CCL-TR-95-004 and
   preprint DESY 95-096 (1995).

\end{thebibliography}

\end{document}